\documentclass{article}

\usepackage{authblk}
\usepackage[authoryear, round]{natbib}
\usepackage[text={7in,8in},centering]{geometry}
\usepackage{geometry}
\usepackage{url}	  
\usepackage{caption}
\usepackage{float}
\usepackage{adjustbox}
\usepackage{multirow}
\usepackage{booktabs}
\usepackage{amsmath, amssymb}
\usepackage{bm}
\usepackage{graphicx}
\usepackage{xcolor}
\usepackage{array}
\usepackage{stfloats} 

\providecommand{\I}{\mathbf{I}}

\providecommand{\x}{\mathbf{x}}
\providecommand{\X}{\mathbf{X}}
\providecommand{\y}{\mathbf{y}}

\providecommand{\zero}{\mathbf{0}}

\providecommand{\lam}{\lambda}

\providecommand{\bb}{\boldsymbol{\beta}}

\providecommand{\bep}{\boldsymbol{\epsilon}}

\providecommand{\bg}{\boldsymbol{\gamma}}

\providecommand{\pr}{\textrm{Pr}}
\providecommand{\Ex}{\mathbb{E}}

\title{Overlapping group logistic regression with applications to genetic pathway selection}
\author{Yaohui Zeng}
\author{Patrick Breheny}
\affil{Department of Biostatistics, University of Iowa, Iowa City, Iowa 52242, U.S.A}

\date{\today}

\begin{document}
\maketitle

\begin{abstract}
Discovering important genes that account for the phenotype of interest has long
been challenging in genomewide expression analysis. Analyses such as Gene Set
Enrichment Analysis (GSEA) that incorporate pathway information have become
widespread in hypothesis testing, but pathway-based approaches have been largely
absent from regression methods due to the challenges of dealing with overlapping
pathways and the resulting lack of available software. The R package
\texttt{grpreg} is widely used to fit group lasso and other group-penalized
regression models; in this study, we develop an extension,
\texttt{grpregOverlap}, to allow for overlapping group structure using the
latent variable approach proposed by \cite{jacob2009group}. We compare this
approach to the ordinary lasso and to GSEA using both simulated and real
data. We find that incorporation of prior pathway information substantially
improves the accuracy of gene expression classifiers, and we shed light on
several ways in which hypothesis-testing approaches such as GSEA differ from
regression approaches with respect to the analysis of pathway data.

\end{abstract}

\section{Introduction}

Since the original proposal of the lasso by \cite{Tibshirani1996}, penalized
regression methods for variable selection in high-dimensional settings have
attracted considerable attention in modern statistical research. These methods
have been extensively studied in theory and widely applied in practice. Most of
the methods focus on selecting individual explanatory variables (or predictors).
In many settings, however, predictors possess a group structure. Incorporating
this grouping information into the modeling process has the potential to improve
both the interpretability and accuracy of the model.

Consider first the linear regression problem with $J$ non-overlapping groups,
\begin{equation} \label{eq:1}
\y = \sum_{j=1}^J \X^j \bb^j + \bep
\end{equation}
where $\y$ is an $n \times 1$ response vector, $\bep \sim N_n(\zero, \sigma^2
\I)$, $\X^j$ is an $n \times K^j$ matrix corresponding to the $j$th group, $K^j$
is the number of elements in group $j$, and $\bb^j$ is the associated
$K^j \times 1$ coefficient vector. In \eqref{eq:1}, we take $\y$ to be centered,
thereby eliminating the need for an intercept. To perform variable selection at
the group level, \cite{Yuan2006} proposed the group lasso estimator, defined as
the value $\bb$ minimizing
\begin{equation} \label{eq:2}
Q(\bb) = L(\bb \mid \y, \X) + \lambda \sum_{j=1}^J \sqrt{K^j} \| \bb^j \|
\end{equation}
where $\|\cdot\|$ is the Euclidean ($l_2$) norm and $L(\bb \mid \y, \X)$ is the
loss function.  For linear regression, the loss function is simply the residual
sum of squares, i.e., $\|\y-\X\bb\|^2/2n$.  For other models, it can be any term that quantifies the fit of the model; for example, \cite{Meier2008}
extended the group lasso selection to logistic regression by using the negative log-likelihood as the loss function. The second term in \eqref{eq:2} is
called the group lasso penalty or $l_1/l_2$ penalty since it is the weighted sum
of the $l_2$ norms of the group coefficient vectors. The group lasso penalty
leads to variable selection at the group level. That is, the coefficient
estimates of the variables in the $j$th group will be all non-zero if group $j$
is selected and all zero otherwise.


An obvious limitation of the group lasso, however, is that it assumes that the
groups do not overlap.  This introduces a barrier to its application for many
problems where variables may be included in more than one group. For instance,
in the analysis of gene expression profiles, individual genes can be grouped
into pathways, in which the collective action of several genes is required in
order for the cell to carry out a complicated function.  These pathways
generally overlap with each other as one gene can play a role in multiple
pathways.

In recent years, various pathway-based approaches have been proposed for
analyzing gene expression data~\citep{nam2008gene}. One key assumption of these
methods is that weak expression changes in individual genes are coordinated and
can be combined in groups to produce stronger signals. Hence, by incorporating
prior pathway information, these approaches aim to identify differentially
expressed pathways, instead of individual genes. Compared to traditional
single-gene tests, pathway-based tests often lead to higher statistical power
and better biological interpretation.  Among the pathway-testing approaches,
Gene Set Enrichment Analysis \citep[GSEA, ][]{mootha2003pgc,
  subramanian2005gene} has been widely used.  The hypothesis testing framework has certain limitations for pathway analysis, however, such as the inability to
account for the effect of multiple genes simultaneously, and it is not
well-suited to using gene expression and pathway data to predict biological
outcomes~\citep{goeman2007analyzing}.

On the other hand, pathway-based approaches have been largely absent from
regression methods due to the challenges of dealing with overlapping pathways in
regression models. To address this issue, \cite{jacob2009group} proposed a
\textit{latent group lasso} approach for variable selection with overlapping
groups, making it possible to perform pathway selection under the general linear modeling framework.

In this paper, we formulate the overlapping group logistic regression model for
pathway selection, and compare this overlapping group lasso (OGLasso) approach
to both the ordinary lasso and GSEA via both simulation and real data
studies. The paper is organized as follows. In Section 2, we review the
overlapping group lasso approach, and construct the overlapping group lasso
model. In addition, we give a brief introduction to GSEA, along with some discussions.
In Section 3, we first compare the ordinary lasso and OGLasso in terms of model
accuracy with simulated data. Then we examine the group selection accuracy of
OGLasso and GSEA under different simulation settings. In addition, we provide
two real data studies in Section 4. We conclude the paper with final discussions
in Section 5.

In addition, we have provided a publicly available implementation of the
overlapping group lasso method described in this article through the R package
\texttt{grpregOverlap}. This package serves as an extension of the R package
\texttt{grpreg}, which provides a variety of functions for fitting penalized
regression models involving grouped predictors, but requires those groups to be
non-overlapping.

\section{Methods}
\subsection{Overlapping group lasso}

Suppose the $p$ predictors $\{x_1, x_2, \ldots, x_p\}$ are assigned
into $J$ possibly overlapping groups (i.e., a given predictor $x_i$ may be
included in more than one group).  The group lasso estimator \eqref{eq:2} does
not necessarily select groups in this overlapping setting.  For example, suppose
$p=3$ and $J=2$, with one covariate shared between the two groups: group ``A''
and group ``B'', with group A truly related to the outcome.  If group B is not
selected, then all of its coefficients are zero, even though one coefficient
also appears in group A.  Thus, group A is only partially selected.  This
problem is greatly exacerbated as the groups grow in size and complexity, and is
described in greater detail in \citet{jenatton2011structured}.

To select entire groups of covariates in the overlapping setting,
\cite{jacob2009group} proposed the overlapping group lasso, formulated as
\begin{align} \label{eq:3}
  \begin{split}
  \min_{\bb} \; Q(\bb) &= L(\bb \mid \y, \X) + \lam \sum_{j=1}^J \sqrt{K^j} \| \bg^j \|\\
  \textrm{subject to } \bb &= \sum_{j=1}^J \bg^j,
  \end{split}
\end{align}
where $\{ \bg^j \}^J_{j=1}$ are $J$ so-called latent coefficient vectors. The
collection of latent vectors $\bg^j = (\gamma_1^j, \gamma_2^j, \ldots,
\gamma_p^j)'$ satisfies $\sum_{j=1}^J \bg^j = \bb$, and $\gamma_k^j = 0$ if
$x_k$ does not belong to group $j$, with $\gamma_k^j \neq 0$ otherwise.

The idea of model~\eqref{eq:3} is to decompose the original coefficient vector
into a sum of group-specific latent effects.  This decomposition allows us to
apply the group lasso penalty to the latent vectors $\{ \bg^j \}^J_{j=1}$, which
do not overlap, instead of the original, overlapping coefficients.
Consequently, when a latent vector $\bg^j$ is selected, all covariates in group
$j$ will be selected, even if some members of the group are also involved in unselected groups.


\begin{figure}[H] 
\centering
\includegraphics[scale=.32]{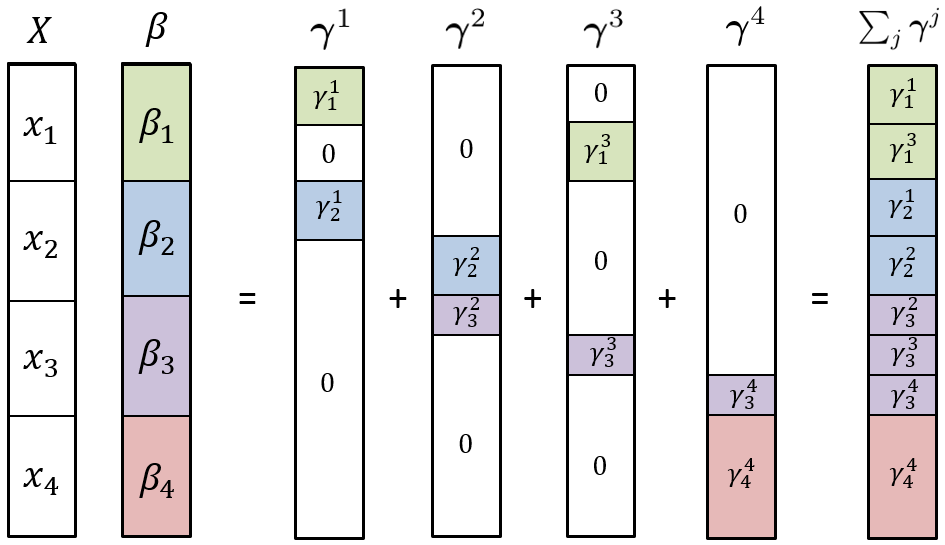}
\caption{The coefficient decomposition of overlapping group lasso.}
\label{fig:1}
\end{figure}

Figure~\ref{fig:1} illustrates the coefficient decomposition mechanism stated
above. Suppose that there are 4 variables $x_1, x_2, x_3, x_4$ that are included
in four groups, $S^1 = \{x_1, x_2\}, \, S^2 = \{x_2, x_3\},\, S^3 = \{x_1,
x_3\}, \, S^4 = \{x_3, x_4\}$, where $S^j$ denote the set of variables in group
$j$.  Since $x_1$ is in both group 1 and 3, $\beta_1$ is thus decomposed into
$\gamma_1^1 + \gamma_1^3$. Likewise, $\beta_3$ is decomposed into
$\gamma_3^2 + \gamma_3^3 + \gamma_3^4$, and so on. Suppose group 1 is the sole
truly nonzero group in this example.  The overlapping group lasso model can
select $\bg^1$, thereby indirectly selecting $\beta_1$ and $\beta_2$ and eliminating
 $\beta_3$ and $\beta_4$ since they do not appear in group 1.  Note that the original
  group lasso cannot accomplish this -- if group 3 is eliminated, then 
  predictor 1 is eliminated as well since it belongs to group 3.

Based on the coefficient decomposition, model~\eqref{eq:3} can be transformed 
into a new minimization problem~\citep{obozinski2011group} with respect to $\bg$:
\begin{equation} \label{eq:latent}
\min_{\bg}  \; Q(\bg) = L(\bg \mid \y,\,\widetilde{\X}) + \lambda \sum_{j=1}^J \sqrt{K^j} \| \bg^j \|.
\end{equation}
Here, $\bg$ in principle consists of all elements of $\bg^j$, although in
practice one can leave off the zero elements as they have no effect on the
objective function.  The new design matrix $\widetilde{\X}$ is constructed by
duplicating the columns of overlapped variables in the raw design matrix $\X$,
where appropriate, to match the elements of $\bg$.  The equivalence of the loss
functions $L(\bb \mid \y, \X)$ and $L(\bg \mid \y,\,\widetilde{\X})$ can be
seen by observing that $\X \bb = \X \sum_j \bg^j = \widetilde{\X} \bg$.

The implication of~\eqref{eq:latent} is that the overlapping group lasso problem
is equivalent to a classical group lasso in an expanded, non-overlapping
space. This is of considerable practical convenience, as it allows us to solve~\eqref{eq:latent} using computationally efficient algorithms that have previously
 been developed for the group lasso \citep{Breheny2015}.


\subsection{Overlapping group logistic regression}

It is relatively straightforward to extend~\eqref{eq:latent} to models other
than linear regression; in this section, we describe its application to
penalized logistic regression in the presence of overlapping groups.  Here, $\y$
is the response vector of binary entries, and the intercept $\beta_0$ cannot be
removed by centering $\y$. For convenience, we assume the first column of the
design matrix $\X$ is the unpenalized column of 1's for the intercept $\beta_0$,
and denote $\x_i = (1, x_{i1}, \ldots, x_{ip})'$ for $i=1, \ldots, n$.
Correspondingly, we denote $\bb = (\beta_0, \beta_1, \ldots, \beta_p)'$.  The
logistic regression model is
\begin{equation} \label{eq:logistic}
\pr (y_i = 1 \mid \x_i) = \pi_i = \frac{\exp(\x_i' \bb)}{1+\exp(\x_i' \bb)}.
\end{equation}
The corresponding loss function is the (scaled) negative log-likelihood
function,
\begin{equation*} 
L(\bb \mid \y, \X) = - \frac{1}{n} \sum_{i=1}^n \{y_i (\x_i' \bb) - \log(1 + \exp(\x_i' \bb)) \}.
\end{equation*}

We can then duplicate the columns of the overlapped covariates, expanding the 
design matrix to $\widetilde{\X}$ as described previously, and construct the 
overlapping group logistic regression model in the same fashion as 
model~\eqref{eq:latent}, with
\begin{equation}
L(\bg \mid \y,\,\widetilde{\X}) = - \frac{1}{n} \sum_{i=1}^n \{y_i (\widetilde{\x}_i' \bg) - \log(1 + \exp(\widetilde{\x}_i' \bg)) \}
\end{equation} 
where $\widetilde{x_i}'$ is the $i$th row of the expanded design matrix 
$\widetilde{\X}$, and the first element of $\bg$ is the unpenalized intercept $\beta_0$.

\subsection{Gene set enrichment analysis (GSEA)}

Among the hypothesis-testing approaches for pathway selection, GSEA stands out
due to its relative simplicity and for preserving the gene-gene dependencies
that occur in real biological data~\citep{tamayo2012limitations}.

The procedure of GSEA~\citep{subramanian2005gene} starts with ranking the $p$
genes by the correlation, $r_j$, between each gene and the phenotype. Then a
test statistic, the enrichment score (ES), is calculated for each gene set by
walking down the ranked gene list and accumulating the correlation information:
increasing ES by $|r_i|^{\alpha} / \sum_{j \in S} |r_j|^{\alpha}$ if gene $i$ is
included in gene set $S$; decreasing ES by $1/(p - |S|)$ otherwise. Here
$\alpha$ is a pre-specified exponent parameter. When $\alpha = 1$, ES
corresponds to the normalized Kolmogorov-Smirnov statistic. Next, the
significance level of the ES is assessed by a permutation test. Finally, the
significance of the gene sets is determined by controlling the false discovery
rate (FDR).


Though widely used, GSEA also has several limitations. First, GSEA may be biased
in favor of larger gene sets by systematically assigning those gene sets higher
ES~\citep{damian2004statistical}; Second, it implicitly assumes genes within the
same gene set show coordinated (i.e., either all positive or all negative)
associations with the phenotype, making it less likely to detect sets in which
the genes are heterogeneous with respect to the direction of association with
the phenotype~\citep{dinu2007improving}.



 
There are inherent differences between GSEA and the proposed overlapping group 
logistic regression method in the sense that GSEA treats the phenotype as fixed
 and gene expression as random, while regression-based methods do the opposite.  
 Thus, GSEA tends to be more appropriate in settings where the phenotype can be 
 directly manipulated by the experiment (e.g., knockout mice), while regression 
 is more appropriate in observational settings (e.g., predicting patient outcomes). 
  Nevertheless, there are many situations in which either method could reasonably 
  be used, and therefore, it is of interest to compare the selection properties of 
  the two approaches.

\section{Simulation studies}

In all of the simulation studies, we use the term ``null group'' to denote a
group whose coefficients are all equal to zero in the true model, and ``true
group'' to denote a group with all non-zero coefficients in the true model. In
addition, we refer to $\| \bg^j \|$ as the effect of group $j$, and $\gamma_k^j$
as the latent effect of covariate $k$ in group $j$.

\subsection{Overlapping group lasso vs. ordinary lasso}
\label{sect:ogl2ol}
We start by comparing the overlapping group lasso (OGLasso) with the ordinary
lasso in terms of estimation and prediction accuracy. We use root mean squared
error (RMSE) to measure estimation accuracy and misclassification error (ME) to
measure prediction accuracy, defined as follows:
\begin{equation*}
\text{RMSE} = \sqrt{\frac{1}{p} \sum_{k=1}^{p} (\beta_k - \hat{\beta}_k)^2} ~ ;  \quad  \text{ME} = \frac{\text{\# incorrectly classified}}{\text{Sample size}}
\end{equation*}
It should be noted that we compute ME based on a new response vector generated 
by the same design matrix for each replication. Specifically, given a design 
matrix $\X$, two response vectors $\y$ and $\y^{*}$ are simulated. The data $\{\X, \y\}$ 
is used to fit the model, and its prediction accuracy is tested on data $\{\X, \y^{*}\}$.

We consider two simulations with different settings described as follows.

\smallskip
\textit{\textbf{Setting 1: Synthetic data.}} We begin with synthetic data 
where there are 15 groups of covariates. All covariate values are simulated 
independently from a standard Gaussian distribution. 
The group sizes and overlap structure are presented below.

\smallskip
\begin{adjustbox}{max width=\linewidth, keepaspectratio, center}
\begin{tabular}{l *{10}{@{\quad}c}}
  ID: & $1 \quad 2$ & $3$ & $4 \quad 5$ & $6$ & $7 \quad 8$ & $9$ & $10 \quad 11$ & $12$ & $13 \quad 14$ & $15$ \\
  Size: & $\underbrace{10 \quad 10}_3$ & $10$ & $\underbrace{10 \quad 10}_3$ & $10$ & 
    $\underbrace{10 \quad 10}_3$ & $10$ & $\underbrace{10 \quad 10}_3$ & $10$ & $\underbrace{10 \quad 10}_3$ & $10$
\end{tabular}
\end{adjustbox}
\smallskip

The number underneath the brace is the number of members shared between those
two groups.  For example, group 1 contains 10 members, as does group 2, but the
two groups contain only 17 unique predictors, as 3 predictors are present in both
groups. As a result, the total dimension in this setting is $p=135$. By design,
groups 1, 4, 7, 10, and 13 are set to be true groups. The sample size is set to be 
$n=50$ to be consistent with that in \textit{\textbf{Setting 2}} as below.

\textit{\textbf{Setting 2: Real data.}} For this simulation, a real gene
expression profile data set in the p53 study~\citep{subramanian2005gene} is used
as the design matrix to mimic the complicated correlation and overlapping
structures in real biomedical applications. This design matrix is fixed for each
independent replication. Here, the sample size $n=50$, the number of genes
$p=4301$, and the number of pathways (groups) is 308; a more detailed
description of the study is given in Section~\ref{realData}.  We chose 5
pathways, with sizes 15, 16, 20, 26, and 40, to represent the true groups in
this simulation.  The number of overlaps between the 5 pathways ranges from 0 to
9.

\smallskip

In both of the two above settings, the true group effect of each of the 5 true
groups is set to be equal, and the latent effects are also set to be equal
within each true group. In this way, the true coefficient vector is uniquely
specified. Then given the design matrix, the responses are generated according
to~(\ref{eq:logistic}) for each independent replication. The true group effect
is varied from 1 to 5 to simulate different magnitudes of signals.

\begin{figure}[H]
\centering
\includegraphics[scale=.5]{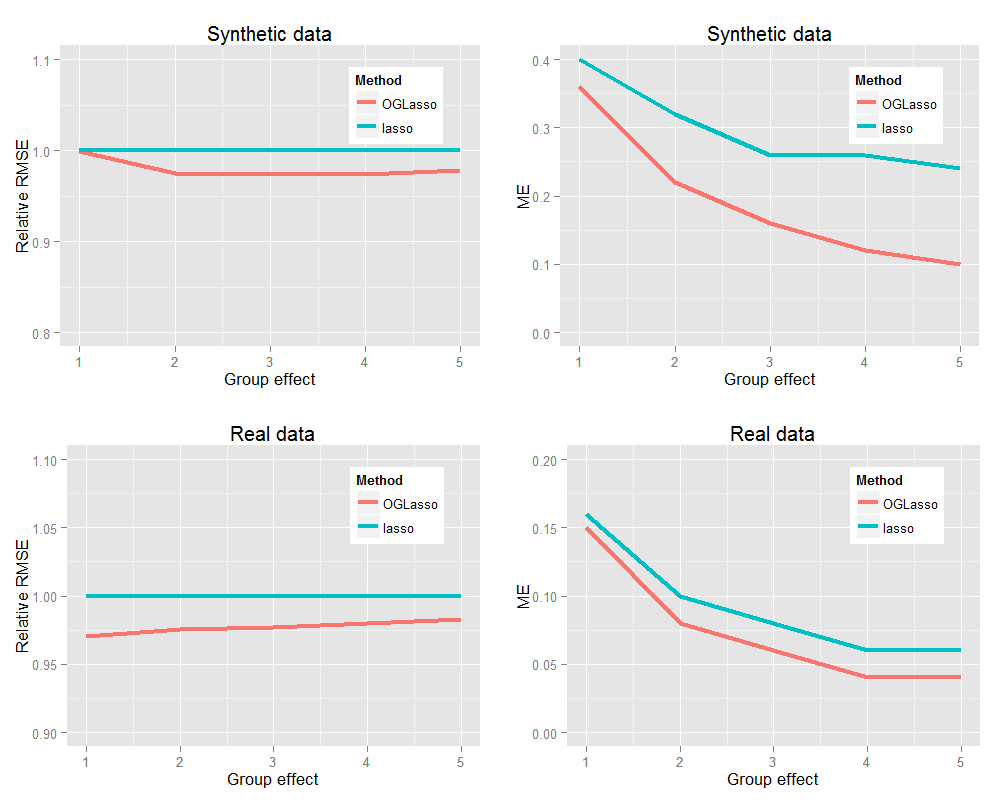}
\caption{Accuracy of OGLasso and ordinary lasso with respect to the magnitude of
  the group effect. Top two panels summarize results for the \textbf{\textit{Synthetic data}} 
  simulation, while bottom two panels are for the \textbf{\textit{Real data}} simulation. 
  Left panels: Median RMSE relative to ordinary lasso over 500 replications. Right panels: 
  Median ME over 500 replications.}
\label{fig:og2ng}
\end{figure}

Figure~\ref{fig:og2ng} illustrates the estimation and prediction accuracy of the
proposed grouped variable selection method, as compared to the ordinary lasso,
for both settings. The top two panels show results for the synthetic data
simulation, while the bottom two panels are for the real data simulation. The
left panels illustrate the median RMSE relative to ordinary lasso over 500
replications, while the right panels compare the methods in terms of ME. OGLasso
consistently achieves a lower median RMSE than that of the lasso in both
synthetic and real data simulations.  As expected, the misclassification error
by both methods decreases as the coefficient magnitude increases.  More
interestingly, the misclassification error by OGLasso can be substantially lower
than that of ordinary lasso. In the synthetic data simulation, for example, the
misclassification error by OGLasso is more than 10\% lower than that of ordinary
lasso when the group effect is 4.  The two methods are more similar in terms of
predictive accuracy on the real data, where the dimensionality is much higher
and correlation structure more complicated.  Nevertheless, the prediction
accuracy can still be improved by around 2\% with OGLassso compared to
ordinary lasso.

\subsection{Overlapping group lasso vs. GSEA}
\label{Sec:oglasso-vs-gsea}

In this section, we use simulated data to compare the selection properties of
the overlapping group lasso against GSEA in a variety of different
settings. Because OGLasso and GSEA do not estimate the same quantities and GSEA
does not produce predictions, the only way to compare them is with respect to
selection accuracy.  To ensure a fair comparison, we use each method to select a
fixed number of groups.  We then evaluate the group selection accuracy by the
true discovery rate (TDR):
\begin{equation*}
\text{TDR} = \frac{\text{\# of true groups selected}}{\text{\# of groups 
selected}},
\end{equation*}
where the \# of groups selected was fixed at 5 (i.e., each method was used to
identify the five most important-looking groups).  In each of the following
simulations, the results are based on sample size $n=100$ and averaged over 500
independent replications.

\smallskip
\textit{\textbf{Setting 3: Unequal group size.}} First, we investigate the
performance of the two approaches when group sizes are unequal. In this
simulation, the design matrix consists of 15 groups with all covariate values
simulated independently from a standard Gaussian distribution. The group sizes
and overlap structure are shown below.

\smallskip
\begin{adjustbox}{max width=\linewidth, keepaspectratio, center}
\begin{tabular}{l *{10}{@{\quad}c}}
  ID: & $1 \quad 2$ & $3$ & $4 \quad 5$ & $6$ & $7 \quad 8$ & $9$ & $10 \quad 11$ & $12$ & $13 \quad 14$ & $15$ \\
  Size: & $\underbrace{3 \quad 3}_1$ & $3$ & $\underbrace{6 \quad 6}_2$ & $6$ & 
    $\underbrace{9 \quad 9}_3$ & $9$ & $\underbrace{15 \quad 15}_5$ & $15$ & $\underbrace{24 \quad 24}_8$ & $24$
\end{tabular}
\end{adjustbox}
\smallskip

The overlap here is designed to be 1/3 of the size of overlapped
groups. As a result, the total dimension in this setting is $p=152$. Moreover,
groups 1, 4, 7, 10, and 13 are set to be true groups with $\| \bg^j \| = 5$, and
the others are null groups with $\bg^j = \zero$.
The latent effects are again set to be equal within each true group.

\begin{table}[H] 
\caption{The mean (standard error) of TDR and average size of selected groups of
 OGLasso and GSEA over 500 replications. }
\begin{adjustbox}{max width=\linewidth, keepaspectratio, center}
\begin{tabular}{lcc}
\toprule
Method & TDR & Average size \\
\hline
OGLasso & 0.77 (0.01) & 8.8 (0.1) \\
GSEA & 0.79 (0.01) & 11.0 (0.1) \\
\bottomrule
\end{tabular}
\end{adjustbox}
\label{tab:GS}
\end{table}

Table~\ref{tab:GS} summarizes the mean TDR and size of selected groups for the
overlapping group Lasso and GSEA over 500 replications. The two methods are
comparable in terms of TDR, while the average size of selected groups from GSEA
is slightly larger.

\begin{figure}[H] 
\centering
\includegraphics[scale = 0.4]{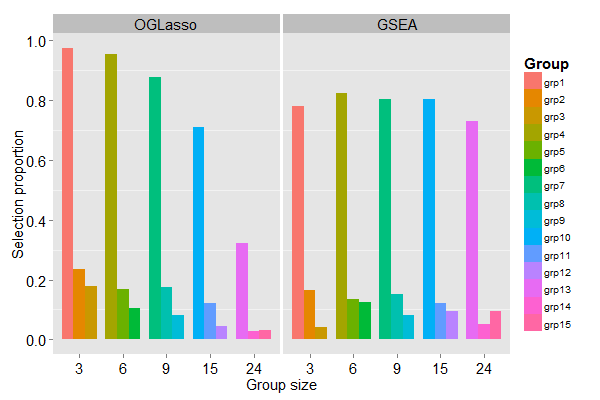}
\caption{Comparison of the proportion of each group being selected over 500 replications.}
\label{fig:GS}
\end{figure}

The proportion of each group selected is depicted in Figure~\ref{fig:GS}.
OGLasso tends to favor groups with smaller size, while GSEA has roughly an equal
probability of selecting a true group regardless of its size.  This is
understandable, as regression-based methods have a built-in mechanism for
encouraging parsimony, unlike GSEA.  Whether this preference for smaller groups
is desirable or not depends on the application and the scientific goals of the
study.

\smallskip
\textit{\textbf{Setting 4: Heterogeneous gene effects.}} Previous studies have
shown that GSEA is less likely to detect sets of genes containing both positive
and negative associations with the phenotype~\citep{dinu2007improving}.  This is
because, by pooling together correlations, GSEA assumes that the genes in a set
have a coordinated effect -- i.e., that they all act in the same direction.  In
this simulation, we examine this aspect of GSEA further and demonstrate that
the exhibition of heterogeneous effects among genes in a set deteriorates the
statistical power of GSEA.

We employ the same configuration as in \textit{\textbf{Setting 1}} of Section
\ref{sect:ogl2ol} for the design matrix (except that the sample size here is
$n=100$), but specify the true coefficient values in a different manner. Specifically, we draw the true latent coefficients $\gamma_k^j$ for each
true group from a Unif($\mu-\sigma, \mu + \sigma$) distribution. Here $\sigma$
is a parameter that controls the degree of heterogeneity (or variability) of the
gene effects.  The larger $\sigma$ is, the more heterogeneous the effects are.
In this simulation, we vary $\sigma$ to examine the effect of heterogeneity on
the TDR of each method.

On a technical note, it must be pointed out that varying $\sigma$ will also
change the group effect, $\| \bg^j \|$. To suppress this possibly confounding
effect, we adjust $\mu$ along with $\sigma$ so that the (root-mean-square) group 
effect remains constant.  Specifically, choosing $\mu = \sqrt{\frac{5}{2} - \frac{1}{3} \sigma^2}$ 
results in a constant $\sqrt{\Ex(\| \bg^j \|^2)} = 5$ for all values of $\sigma$.


\begin{figure}[H]
\centering
\includegraphics[scale=.4]{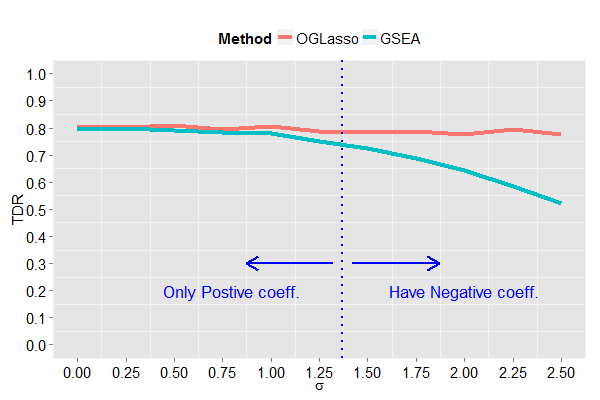}
\caption{Comparison of OGLasso and GSEA in terms of TDR as a function of
  heterogeneity parameter $\sigma$. The blue dotted line indicates $\sigma =
  1.37$, after which negative coefficients can occur by design. The mean values
  over 500 replications are displayed.}
\label{fig:hetero}
\end{figure}

Figure~\ref{fig:hetero} compares OGLasso and GSEA in terms of TDR as a function
of $\sigma$. OGLasso is essentially unaffected by heterogeneity: it detects
approximately 4 out of the 5 true groups regardless of the magnitude of
heterogeneity.  In contrast, the TDR of GSEA decreases as $\sigma$
increases. This effect is apparent even when all genes in a group have a consistent 
direction, although the effect is much more significant for $\sigma > 1.37$, at which
 point it is possible for genes within a true group to have opposite directions.

\smallskip
\textbf{\textit{Setting 5: Correlation among genes.}} In this simulation, we
assess how correlation among genes affects group selection. We use the same
settings for the groups and overlap structure as in \textbf{\textit{Setting 1}},
where $p = 135$. The true coefficients are fixed so that the group effect $\|
\bg^j \| = 5$ for each true group, and that all latent effects $\gamma_k^j$
within a true group $j$ are equal. In this setting, covariates are no longer
independent, but are instead simulated from a multivariate Gaussian distribution
with mean $\zero$ and variance $\boldsymbol \Sigma$.  We impose a block-diagonal
covariance structure with 5 compound-symmetric blocks, as shown below:

\begin{equation*}
\boldsymbol \Sigma = \begin{bmatrix}
\Sigma & & & & \\
& \Sigma & & & \\
& & \Sigma & & \\
& & & \Sigma & \\
& & & & \Sigma
\end{bmatrix},
\text{where } \Sigma = \begin{bmatrix}
1 & \rho & \cdots & \rho \\
\rho & 1 & \cdots & \rho \\
\vdots & \vdots & \ddots & \vdots \\
\rho & \rho & \cdots & 1 \\
\end{bmatrix}_{27 \times 27}
\end{equation*}

\begin{figure}[H]
\centering
\includegraphics[scale=.4]{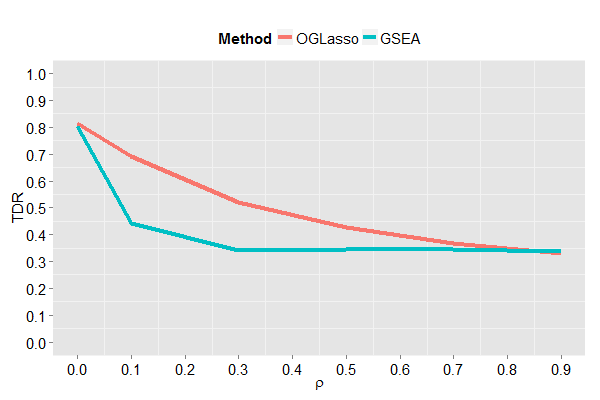}
\caption{Comparison of OGLasso and GSEA in terms of TDR as a function of pairwise
 correlation $\rho$. The mean values over 500 replications are displayed.}
\label{fig:corr}
\end{figure}

Figure~\ref{fig:corr} compares OGLasso and GSEA in terms of TDR as a function of
pairwise correlation $\rho$. As expected, TDR of both methods decreases as the
correlation among genes increases. However, GSEA is much more strongly affected
by correlation than the overlapping group lasso. For example, as $\rho$
increases from 0 to 0.1, the TDR of GSEA drops from around 0.8 to 0.45. This 
ability -- to adjust for correlation between pathways -- is one of the primary 
potential advantages of a regression-based approach over a hypothesis-testing 
approach, which is limited to considering a single pathway at a time.
 
\section{Real data studies} \label{realData}

In this section, we analyze the data from two gene expression studies reported 
in~\cite{subramanian2005gene}, one involving the mutational status of p53 in cell 
lines and the other involving the prognosis of lung cancer patients.

The p53 study aims to identify pathways that correlated with the mutational
status of the gene p53, which regulates gene expression in response to various
signals of cellular stress. The p53 data ~\citep{olivier2002iarc} consist of 
50 cell lines, 17 of which are
classified as normal and 33 of which carry mutations in the p53 gene.  To be
consistent with the analysis in~\cite{subramanian2005gene}, 308 gene sets that
have size between 15 and 500 are included in our analysis.  These gene sets
contain a total of 4301 genes.

The lung cancer data~\citep{beer2002gene} contains gene expression profiles in
86 tumor samples, out of which 24 are classified as ``poor'' outcome and the
remaining as ``good'' outcome. The data sets are preprocessed in the same
fashion as in the p53 study, resulting in 258 gene sets that contain a total of
3256 genes. Compared to the p53 data, the lung cancer data show much weaker
signals: no individual gene is found to be significant in a conventional
single-gene analysis.

We first compare the OGLasso to the ordinary lasso in terms of
prediction accuracy.  For each method, 10-fold cross-validation was used to
choose the regularization parameter $\lambda$.

\begin{table} [H]
\caption{Real data studies: 10 fold cross-validated misclassification error (ME)
  for different models. ``Baseline'' is the intercept-only model.}
\begin{adjustbox}{max width=\linewidth, keepaspectratio, center}
\begin{tabular}{lcc}
\toprule
Method & p53 study & lung cancer study \\
\hline
Baseline & 0.34 & 0.28 \\
lasso & 0.26 & 0.30 \\
OGLasso & 0.20 & 0.27 \\
\bottomrule
\end{tabular}
\end{adjustbox}
\label{realdata1}
\end{table}

Indeed, as shown in Table~\ref{realdata1}, the incorporation of pathway
information into the regression model produces more accurate predictions in both
studies. In the p53 study, where the signals are relatively strong, the
misclassification error of the ordinary lasso is 8\% lower than that of the
intercept-only model.  The OGLasso, however, can further lower
the error by an additional 6\%. In the lung cancer study, due to a small
signal-to-noise ratio, the ordinary lasso performed even worse than the
intercept-only model. The OGLasso, however, was able to improve on the predictions 
of the intercept-only model, albeit only slightly.

We now turn to comparing the pathways selected by OGLasso and GSEA. Again,
10-fold cross-validation is used to select $\lambda$ for OGLasso, while a FDR
cutoff of 0.25 was used to select pathways with GSEA.  Table~\ref{realdata2}
lists the number of pathways, the number of total genes and number of unique
genes in those selected pathways by OGLasso and GSEA.  In both studies, GSEA
selects more pathways than OGLasso, especially in the lung cancer study (21
vs. 3). Moreover, in agreement with our earlier simulation results, GSEA selects
substantially larger pathways than OGlasso.  For example, in the lung cancer
study the average pathway size for GSEA is 820/21=39 genes, while the average
size for OGLasso is only 51/3=17 genes.

\begin{table}[H]
\caption{Real data studies: number of selected pathways (\# Pathways), number of total
 genes (\# Total Genes), and number of unique genes (\# Unique Genes) in selected 
 pathways by OGLasso and GSEA.}
\begin{adjustbox}{max width=\linewidth, keepaspectratio, center}
\begin{tabular}{lccccccc}
\toprule
\multirow{3}[3]{*}{Method} & \multicolumn{3}{c}{p53 study} &  \multicolumn{3}{c}{lung cancer study} \\
\cmidrule(lr){2-4} \cmidrule(lr){5-7}
 & \multicolumn{1}{c}{\# Pathways}
 & \multicolumn{1}{c}{\# Total}
  & \multicolumn{1}{c}{\# Unique}
 & \multicolumn{1}{c}{\# Pathways}
 & \multicolumn{1}{c}{\# Total}
  & \multicolumn{1}{c}{\# Unique} \\
   & \multicolumn{1}{c}{}
 & \multicolumn{1}{r}{Genes}
  & \multicolumn{1}{c}{Genes}
 & \multicolumn{1}{c}{}
 & \multicolumn{1}{c}{Genes}  
  & \multicolumn{1}{c}{Genes} \\
\midrule
OGLasso & 3 & 46  & 44 & 3 & 51 & 50 \\
GSEA    & 6 & 139 & 105 & 21 & 820 & 629  \\
\bottomrule
\end{tabular}
\end{adjustbox}
\label{realdata2}
\end{table}

Table~\ref{p53pathway} presents a summary of pathway selection results in the
p53 study that sheds light on the nature of the pathways selected by each
approach.  Naturally, both approaches identify the ``p53Pathway'' as being
associated with p53 mutation status.  However, GSEA also selects pathways
``radiation\_sensitivity'', which shares 9 genes with ``p53Pathway'',
``p53hypoxiaPathway'' (7 shared genes), and ``P53\_UP'' (5 shared genes).  From
 a regression perspective, these four pathways are largely redundant, and the 
 three unselected pathways carry no additional useful information beyond that 
 already contained in the p53 pathway.  On the other hand, OGLasso selects one 
 pathway, ``ck1Pathway'', not identified by GSEA.  Although the ck1 pathway has 
 a weaker marginal relationship with p53 mutation status than the hsp27 and p53 
 pathways, the information it contains is largely independent of the other pathways 
 included in the model (no overlaps with the hsp27 and p53 pathways), potentially 
 shedding light on novel p53 relationships that would not be apparent from the 
 GSEA approach.

\begin{table} [H]
\caption{The p53 study: pathways selected by OGLasso and GSEA with FDR  $\leq 0.25$. }
\begin{adjustbox}{max width=\linewidth, keepaspectratio, center}
\begin{tabular}{rcccc}
\toprule
Pathway label & Size & FDR $q$ value & GSEA & OGLasso \\
\hline
hsp27Pathway & 15 & $< .001$ & \checkmark & \checkmark \\
p53hypoxiaPathway & 20 & $< .001$ & \checkmark &       \\
p53Pathway & 16 & $< .001$ & \checkmark & \checkmark   \\
radiation\_sensitivity & 26 & $0.078$ & \checkmark &   \\
P53\_UP & 40 & $0.013$ & \checkmark  \\
rasPathway & 22 & $0.171$ &\checkmark & \\
ck1Pathway & 15 & $0.500$ &  & \checkmark \\
\bottomrule
\end{tabular}
\end{adjustbox}
\label{p53pathway}
\end{table}

The biological interpretations of the pathways selected in the lung cancer study
are less clear due to the weaker signals and more complicated biological
outcome.  Nevertheless, there are some interesting similarities and differences
here as well.  Of the three gene sets selected by OGLasso, one pathway
(ceramide) is also selected by GSEA.  The other two gene sets, although not
selected by GSEA, contain a fair amount of overlap with GSEA-selected sets.  For
example, OGLasso selects the Fas pathway while GSEA selects the p53 pathway.
However, both pathways are involved in apoptosis, and 6 genes are shared between
the two pathways.  The simulation studies of Section~\ref{Sec:oglasso-vs-gsea} 
suggest that differences in the size, heterogeneity, or correlation patterns of
 these pathways provide an explanation for why OGLasso prefers the Fas pathway to the p53 pathway.



\section{Discussion}

Pathway-based approaches for analyzing gene expression data have become
increasingly popular in recent years.  Most methods have approached the problem
from a multiple hypotheses testing perspective.  However, the overlapping group
lasso approach proposed by \cite{jacob2009group} allows the incorporation of
pathway information into regression models as well.  

In this paper, we present evidence that the incorporation of pathway information
can substantially improve the accuracy of gene expression classifiers.
Furthermore, we provide open-source software, publicly available at
\url{cran.r-project.org}, for fitting the overlapping group lasso models
described in this paper.

Finally, this paper provides, to our knowledge, the only systematic comparison of 
overlapping group lasso methods with the GSEA approach.  There is a fundamental 
difference between the two methods: GSEA carries out independent tests of each 
gene set, while the overlapping group lasso is a regression method that considers 
the effect of all pathways simultaneously.  We show that, while there is broad 
agreement between the two, substantial differences between the approaches may arise
 with respect to pathway size, heterogeneity of gene effects, and correlations 
 between gene sets.  These factors, along with the goals and design of the study, 
 should be carefully considered when deciding upon an approach to data analysis.



\bibliographystyle{plainnat} 

\begin{thebibliography}{16}
\providecommand{\natexlab}[1]{#1}
\providecommand{\url}[1]{\texttt{#1}}
\expandafter\ifx\csname urlstyle\endcsname\relax
  \providecommand{\doi}[1]{doi: #1}\else
  \providecommand{\doi}{doi: \begingroup \urlstyle{rm}\Url}\fi

\bibitem[Beer et~al.(2002)Beer, Kardia, Huang, Giordano, Levin, Misek, Lin,
  Chen, Gharib, Thomas, et~al.]{beer2002gene}
David~G Beer, Sharon~LR Kardia, Chiang-Ching Huang, Thomas~J Giordano, Albert~M
  Levin, David~E Misek, Lin Lin, Guoan Chen, Tarek~G Gharib, Dafydd~G Thomas,
  et~al.
\newblock Gene-expression profiles predict survival of patients with lung
  adenocarcinoma.
\newblock \emph{Nature medicine}, 8\penalty0 (8):\penalty0 816--824, 2002.

\bibitem[Breheny and Huang(2015)]{Breheny2015}
Patrick Breheny and Jian Huang.
\newblock Group descent algorithms for nonconvex penalized linear and logistic
  regression models with grouped predictors.
\newblock \emph{Statistics and Computing}, 25\penalty0 (2):\penalty0 173--187,
  2015.
\newblock ISSN 0960-3174.
\newblock \doi{10.1007/s11222-013-9424-2}.
\newblock URL \url{http://dx.doi.org/10.1007/s11222-013-9424-2}.

\bibitem[Damian and Gorfine(2004)]{damian2004statistical}
Doris Damian and Malka Gorfine.
\newblock Statistical concerns about the gsea procedure.
\newblock \emph{Nature genetics}, 36\penalty0 (7):\penalty0 663--663, 2004.

\bibitem[Dinu et~al.(2007)Dinu, Potter, Mueller, Liu, Adewale, Jhangri,
  Einecke, Famulski, Halloran, and Yasui]{dinu2007improving}
Irina Dinu, John~D Potter, Thomas Mueller, Qi~Liu, Adeniyi~J Adewale, Gian~S
  Jhangri, Gunilla Einecke, Konrad~S Famulski, Philip Halloran, and Yutaka
  Yasui.
\newblock Improving gene set analysis of microarray data by sam-gs.
\newblock \emph{BMC bioinformatics}, 8\penalty0 (1):\penalty0 242, 2007.

\bibitem[Goeman and B{\"u}hlmann(2007)]{goeman2007analyzing}
Jelle~J Goeman and Peter B{\"u}hlmann.
\newblock Analyzing gene expression data in terms of gene sets: methodological
  issues.
\newblock \emph{Bioinformatics}, 23\penalty0 (8):\penalty0 980--987, 2007.

\bibitem[Jacob et~al.(2009)Jacob, Obozinski, and Vert]{jacob2009group}
Laurent Jacob, Guillaume Obozinski, and Jean-Philippe Vert.
\newblock Group lasso with overlap and graph lasso.
\newblock In \emph{Proceedings of the 26th Annual International Conference on
  Machine Learning}, pages 433--440. ACM, 2009.

\bibitem[Jenatton et~al.(2011)Jenatton, Audibert, and
  Bach]{jenatton2011structured}
Rodolphe Jenatton, Jean-Yves Audibert, and Francis Bach.
\newblock Structured variable selection with sparsity-inducing norms.
\newblock \emph{The Journal of Machine Learning Research}, 12:\penalty0
  2777--2824, 2011.

\bibitem[Meier et~al.(2008)Meier, Geer, and B�hlmann]{Meier2008}
Lukas Meier, Sara van~de Geer, and Peter B�hlmann.
\newblock The group lasso for logistic regression.
\newblock \emph{Journal of the Royal Statistical Society. Series B (Statistical
  Methodology)}, 70\penalty0 (1):\penalty0 pp. 53--71, 2008.
\newblock ISSN 13697412.
\newblock URL \url{http://www.jstor.org/stable/20203811}.

\bibitem[Mootha et~al.(2003)Mootha, Lindgren, Eriksson, Subramanian, Sihag,
  Lehar, Puigserver, Carlsson, Ridderstr{$\aa$}le, Laurila,
  et~al.]{mootha2003pgc}
Vamsi~K Mootha, Cecilia~M Lindgren, Karl-Fredrik Eriksson, Aravind Subramanian,
  Smita Sihag, Joseph Lehar, Pere Puigserver, Emma Carlsson, Martin
  Ridderstr{$\aa$}le, Esa Laurila, et~al.
\newblock Pgc-1{$\alpha$}-responsive genes involved in oxidative
  phosphorylation are coordinately downregulated in human diabetes.
\newblock \emph{Nature genetics}, 34\penalty0 (3):\penalty0 267--273, 2003.

\bibitem[Nam and Kim(2008)]{nam2008gene}
Dougu Nam and Seon-Young Kim.
\newblock Gene-set approach for expression pattern analysis.
\newblock \emph{Briefings in bioinformatics}, 9\penalty0 (3):\penalty0
  189--197, 2008.

\bibitem[Obozinski et~al.(2011)Obozinski, Jacob, and Vert]{obozinski2011group}
Guillaume Obozinski, Laurent Jacob, and Jean-Philippe Vert.
\newblock Group lasso with overlaps: the latent group lasso approach.
\newblock \emph{arXiv preprint arXiv:1110.0413}, 2011.

\bibitem[Olivier et~al.(2002)Olivier, Eeles, Hollstein, Khan, Harris, and
  Hainaut]{olivier2002iarc}
Magali Olivier, Ros Eeles, Monica Hollstein, Mohammed~A Khan, Curtis~C Harris,
  and Pierre Hainaut.
\newblock The iarc tp53 database: new online mutation analysis and
  recommendations to users.
\newblock \emph{Human mutation}, 19\penalty0 (6):\penalty0 607--614, 2002.

\bibitem[Subramanian et~al.(2005)Subramanian, Tamayo, Mootha, Mukherjee, Ebert,
  Gillette, Paulovich, Pomeroy, Golub, Lander, et~al.]{subramanian2005gene}
Aravind Subramanian, Pablo Tamayo, Vamsi~K Mootha, Sayan Mukherjee, Benjamin~L
  Ebert, Michael~A Gillette, Amanda Paulovich, Scott~L Pomeroy, Todd~R Golub,
  Eric~S Lander, et~al.
\newblock Gene set enrichment analysis: a knowledge-based approach for
  interpreting genome-wide expression profiles.
\newblock \emph{Proceedings of the National Academy of Sciences of the United
  States of America}, 102\penalty0 (43):\penalty0 15545--15550, 2005.

\bibitem[Tamayo et~al.(2012)Tamayo, Steinhardt, Liberzon, and
  Mesirov]{tamayo2012limitations}
Pablo Tamayo, George Steinhardt, Arthur Liberzon, and Jill~P Mesirov.
\newblock The limitations of simple gene set enrichment analysis assuming gene
  independence.
\newblock \emph{Statistical methods in medical research}, page
  0962280212460441, 2012.

\bibitem[Tibshirani(1996)]{Tibshirani1996}
Robert Tibshirani.
\newblock Regression shrinkage and selection via the lasso.
\newblock \emph{Journal of the Royal Statistical Society. Series B
  (Methodological)}, 58\penalty0 (1):\penalty0 pp. 267--288, 1996.
\newblock ISSN 00359246.
\newblock URL \url{http://www.jstor.org/stable/2346178}.

\bibitem[Yuan and Lin(2006)]{Yuan2006}
Ming Yuan and Yi~Lin.
\newblock Model selection and estimation in regression with grouped variables.
\newblock \emph{Journal of the Royal Statistical Society: Series B (Statistical
  Methodology)}, 68\penalty0 (1):\penalty0 49--67, 2006.

\end{thebibliography}

\end{document}